\documentclass[letterpaper,twocolumn,10pt]{article}
\usepackage{usenix-2020-09}

\renewcommand{\textrightarrow}{$\rightarrow$}

\usepackage{tikz}
\usepackage{xcolor}
\usepackage{xspace}
\usepackage{tikz}
\usepackage{amsmath}
\usepackage{pifont}
\usepackage[english]{babel}
\usepackage{blindtext}
\usepackage{lipsum}
\usepackage{amsthm}
\usepackage[ruled,vlined,linesnumbered]{algorithm2e}
\usepackage{booktabs}
\usepackage{amssymb}
\usepackage{amsfonts}

\usepackage{filecontents}
\usepackage{xspace}
\usepackage{subfig}
\usepackage{xcolor}
\usepackage{hhline}
\usepackage{multirow}
\usepackage{soul}
\usepackage{amsthm}

\usepackage{array}
\usepackage{comment}
\usepackage{listings}
\usepackage{dsfont}
\usepackage[title]{appendix}

\usepackage{caption}
\usepackage{ltablex}

\usepackage{ifluatex}

\begin{document}

\newcommand*{\affmark}[1][*]{\textsuperscript{#1}}
\newcommand*{\affaddr}[1]{#1}
\title{\large Towards More Economical Context-Augmented LLM Generation by Reusing Stored KV Cache}

\author{ Hanchen Li$^*$ (Student),
Yuhan Liu$^*$ (Student),
Yihua Cheng (Student),
Kuntai Du (Student),
Junchen Jiang 
\\
\textit{\{lihanc2002,\xspace yuhanl,\xspace yihua98,\xspace kuntai,\xspace junchenj\}@uchicago.edu}
\\
The University of Chicago
}

\newcommand\blfootnote[1]{%
  \begingroup
  \renewcommand\thefootnote{}\footnote{#1}%
  \addtocounter{footnote}{-1}%
  \endgroup
}
\newcommand{\shan}[1]{}
\newcommand{\hank}[1]{ }
\newcommand{\mm}[1]{ }
\newcommand{\cc}[1]{ }
\newcommand{\jc}[1]{ }
\newcommand{\yh}[1]{ }
\newcommand{\hc}[1]{ }

\newcommand{\kt}[1]{ }

\newcommand{\edit}[1]{{\color{black}{#1}}}

\newcommand{\Decision}{Decision\xspace}
\newcommand{\x}{\ensuremath{\mathbf{x}}\xspace}
\newcommand{\y}{\ensuremath{\mathbf{y}}\xspace}
\newcommand{\z}{\ensuremath{z}\xspace}
\newcommand{\X}{\ensuremath{\mathbf{X}}\xspace}
\newcommand{\TestNum}{\ensuremath{T}\xspace}
\newcommand{\APIcomp}{App \circ Filter\xspace}
\newcommand{\Filter}{F\xspace}
\newcommand{\DNN}{DNN\xspace}
\newcommand{\API}{API\xspace}
\newcommand{\Appdecision}{App\xspace}
\newcommand{\total}{\ensuremath{t}\xspace}
\newcommand{\NTarget}{\ensuremath{N}\xspace}
\newcommand{\other}{\ensuremath{O}\xspace}

\newcommand{\Rate}{IDR\xspace}

\newcommand{\T}{\ensuremath{T}\xspace}
\newcommand{\Match}{\ensuremath{M}\xspace}
\newcommand{\BeforeBreak}{\ensuremath{B}\xspace}

\newcommand{\App}{$A$\xspace}
\newcommand{\M}{\ensuremath{M}\xspace}
\newcommand{\F}{\ensuremath{\alpha}\xspace}
\newcommand{\G}{\ensuremath{\beta}\xspace}
\newcommand{\Q}{\ensuremath{\gamma}\xspace}
\newcommand{\ClassId}{\ensuremath{c}\xspace}
\newcommand{\LabelId}{\ensuremath{l}\xspace}
\newcommand{\InputId}{\ensuremath{i}\xspace}
\newcommand{\TraverseId}{\ensuremath{j}\xspace}
\newcommand{\score}{effective score\xspace}
\newcommand{\scores}{effective scores\xspace}
\newcommand{\Tg}{\ensuremath{TC}\xspace}
\newcommand{\TargetSet}{\ensuremath{G}\xspace}
\newcommand{\MaxScore}{\ensuremath{f(\y)}\xspace}

\newcommand{\summary}{decision-process summary\xspace}
\newcommand{\Summary}{Decision-process summary\xspace}
\newcommand{\summaries}{decision-process summaries\xspace}
\newcommand{\target}{target class\xspace}
\newcommand{\targets}{target classes\xspace}
\newcommand{\Target}{Target class\xspace}
\newcommand{\process}{software decision process\xspace}
\newcommand{\Process}{Software decision process\xspace}
\newcommand{\bestapi}{Best-of-all API*\xspace}
\newcommand{\prespec}{Categorized models\xspace}

\newcommand{\genmodel}{generic model\xspace}
\newcommand{\genmodels}{generic models\xspace}
\newcommand{\Genmodel}{Generic model\xspace}
\newcommand{\Genmodels}{Generic models\xspace}

\newcommand{\name}{ChameleonAPI\xspace}
\newcommand{\modeld}{$\textrm{ChameleonAPI}_{basic}$\xspace}
\newcommand{\tool}{ChameleonAPI\xspace}
\newcommand{\toolbasic}{\ensuremath{\textrm{ChameleonAPI}_{basic}}\xspace}
\newcommand{\code}[1]{{\texttt{#1}}}
\newcommand{\term}[1]{\textsf{#1}}
\newcommand{\wl}{\emph{whitelist}\xspace}
\newcommand{\wls}{\emph{whitelists}\xspace}
\newcommand{\ift}{\emph{if/then branch}\xspace}
\newcommand{\ifts}{\emph{if/then branches}\xspace}

\newcommand{\fillme}{{\bf XXX}\xspace}

\newcommand*\circled[1]{\tikz[baseline=(char.base)]{
            \node[shape=circle,fill,inner sep=2pt] (char) {\textcolor{white}{\footnotesize{#1}}};}}

\newcounter{packednmbr}
\newenvironment{packedenumerate}{\begin{list}{\thepackednmbr.}{\usecounter{packednmbr}\setlength{\itemsep}{0.5pt}\addtolength{\labelwidth}{-4pt}\setlength{\leftmargin}{2ex}\setlength{\listparindent}{\parindent}\setlength{\parsep}{1pt}\setlength{\topsep}{0pt}}}{\end{list}}
\newenvironment{packeditemize}{\begin{list}{$\bullet$}{\setlength{\itemsep}{0.5pt}\addtolength{\labelwidth}{-4pt}\setlength{\leftmargin}{2ex}\setlength{\listparindent}{\parindent}\setlength{\parsep}{1pt}\setlength{\topsep}{2pt}}}{\end{list}}
\newenvironment{packedpackeditemize}{\begin{list}{$\bullet$}{\setlength{\itemsep}{0.5pt}\addtolength{\labelwidth}{-4pt}\setlength{\leftmargin}{\labelwidth}\setlength{\listparindent}{\parindent}\setlength{\parsep}{1pt}\setlength{\topsep}{0pt}}}{\end{list}}
\newenvironment{packedtrivlist}{\begin{list}{\setlength{\itemsep}{0.2pt}\addtolength{\labelwidth}{-4pt}\setlength{\leftmargin}{\labelwidth}\setlength{\listparindent}{\parindent}\setlength{\parsep}{1pt}\setlength{\topsep}{0pt}}}{\end{list}}
\let\enumerate\packedenumerate
\let\endenumerate\endpackedenumerate
\let\itemize\packeditemize
\let\enditemize\endpackeditemize

\newcommand{\tightcaption}[1]{\vspace{-0.15cm}\caption{{\normalfont{\textit{{#1}}}}}\vspace{-0.3cm}}
\newcommand{\tightsection}[1]{\vspace{-0.2cm}\section{#1}\vspace{-0.1cm}}
\newcommand{\tightsectionstar}[1]{\vspace{-0.17cm}\section*{#1}\vspace{-0.08cm}}
\newcommand{\tightsubsection}[1]{\vspace{-0.25cm}\subsection{#1}\vspace{-0.1cm}}
\newcommand{\tightsubsubsection}[1]{\vspace{-0.01in}\subsubsection{#1}\vspace{-0.01cm}}

\newcommand{\eg}{{\it e.g.,}\xspace}
\newcommand{\ie}{{\it i.e.,}\xspace}
\newcommand{\etal}{{\it et.~al}\xspace}
\newcommand{\bigO}{\mathrm{O}}
\newcommand{\twlog}{w.l.o.g.\xspac}

\newcommand{\myparashort}[1]{\vspace{0.05cm}\noindent{\bf {#1}}~}
\newcommand{\mypara}[1]{\vspace{0.05cm}\noindent{\bf {#1}:}~}
\newcommand{\myparatight}[1]{\vspace{0.02cm}\noindent{\bf {#1}:}~}
\newcommand{\myparaq}[1]{\smallskip\noindent{\bf {#1}?}~}
\newcommand{\myparaittight}[1]{\smallskip\noindent{\emph {#1}:}~}
\newcommand{\question}[1]{\smallskip\noindent{\emph{Q:~#1}}\smallskip}
\newcommand{\myparaqtight}[1]{\smallskip\noindent{\bf {#1}}~}

\newcommand{\cmark}{\ding{51}}%
\newcommand{\xmark}{\ding{55}}%



\definecolor{backcolour}{rgb}{0.96,0.96,0.96}
\definecolor{codegray}{rgb}{0.5,0.5,0.5}
\definecolor{deepblue}{rgb}{0,0,0.6}
\definecolor{deepred}{rgb}{0.6,0,0}
\definecolor{deepgreen}{rgb}{0,0.5,0}
\lstdefinestyle{mystyle}{
    backgroundcolor=\color{backcolour},   
    commentstyle=\color{codegreen},
    morekeywords={self, True},
    keywordstyle=\color{deepblue},
    numberstyle=\tiny\color{codegray},
    emph={MyClass,__init__,EncodingType,Image},
    emphstyle=\color{deepred},
    stringstyle=\color{deepgreen},
    basicstyle=\ttfamily\footnotesize,
    breakatwhitespace=false,         
    breaklines=true,                 
    captionpos=b,                    
    keepspaces=true,                 
    numbers=left,                    
    numbersep=5pt,                  
    showspaces=false,                
    showstringspaces=false,
    showtabs=false,                  
    tabsize=1
}

\maketitle

\begin{abstract}
Across large language model (LLM) applications, we observe an emerging trend for reusing KV caches to save the prefill delays of processing repeated input texts in different LLM inputs. 
This has led to a broad design space, including colocating stored KV caches with (or close to) GPUs to various KV cache compression.
However, a key question remains unanswered: {\em can these delay reductions also be economically favorable?}
Specifically, we ask whether a developer can use public cloud services to store precomputed KV caches and reuse them to save delay {\em without} incurring more costs in terms of compute, storage, and network.
To answer this question, we propose an validated analytical model for the cloud cost (in compute, storage, and network) of storing and reusing KV caches based on various workload parameters, such as reuse frequency, generated text lengths, model sizes, etc.
Preliminary results show that KV cache reusing is able to save both delay and cloud cost across a range of workloads with long context. And we call more efforts on building more economical context augmented LLM by KV cache reusing. 

\blfootnote{$^*$Equal Contribution}
\end{abstract}

\tightsection{Introduction}


Nowadays, a wide range of applications (\eg document-based question answering) often use large language models (LLMs) after prepending the LLM inputs with {\em long contexts} (\eg related documents) that may contain thousands of tokens or more. 
In this paper, we refer such long prefixes as contexts.

Prior work found that different user requests may need to use the same context~\cite{joshi2017triviaqa}. Therefore, the computation results of the context (in the format of KV caches), once computed and cached in the GPU, can be reused across those jobs that use the same context~\cite{vllm,chunkattention}. For example, two requests, ``What was the highest source of revenue of Apple in the last year?'' and ``Which supplier for Apple offers the lowest average price?'', may need to use the same annual fiscal report as the context.


However, the design challenge of such KV cache reusing scheme, is \textit{where to store the pre-computed KV cache of the context}?
One potential design is to store the KV cache in the GPU~\cite{vllm, chunkattention}, which is feasible when there are only limited number of contexts that the user will access (\eg a dedicated server that answers the question about a specific software), but is less favorable when a wide range of context may be accessed, as the GPU will not have sufficient GPU memory to store the KV cache of all the contexts.

An alternative solution is to store the KV cache of the context in a dedicated storage server and transfer the KV cache into GPU memory when new jobs requesting contexts arrive.
Acknowledging that such an approach incurs extra transfer delay and storage cost, we argue that such an approach is favorable compared with text recomputation in terms of both end-to-end delay and economic cost.
 By building an analytical model for developers to compare service costs given their workload pattern and cloud service pricing policy, we show that reusing KV cache from storage is more economical than text recomputation. Moreover, we validate this result by simulation under various workloads. This proves that the end-to-end delay and economic cost of serving LLM inference jobs with KV cache are consistently lower than without storing KV cache.
We hope this poster sparks more discussions toward realizing the full potential of such systems.



\begin{figure}[]
\centering
\includegraphics[width=0.99\linewidth]{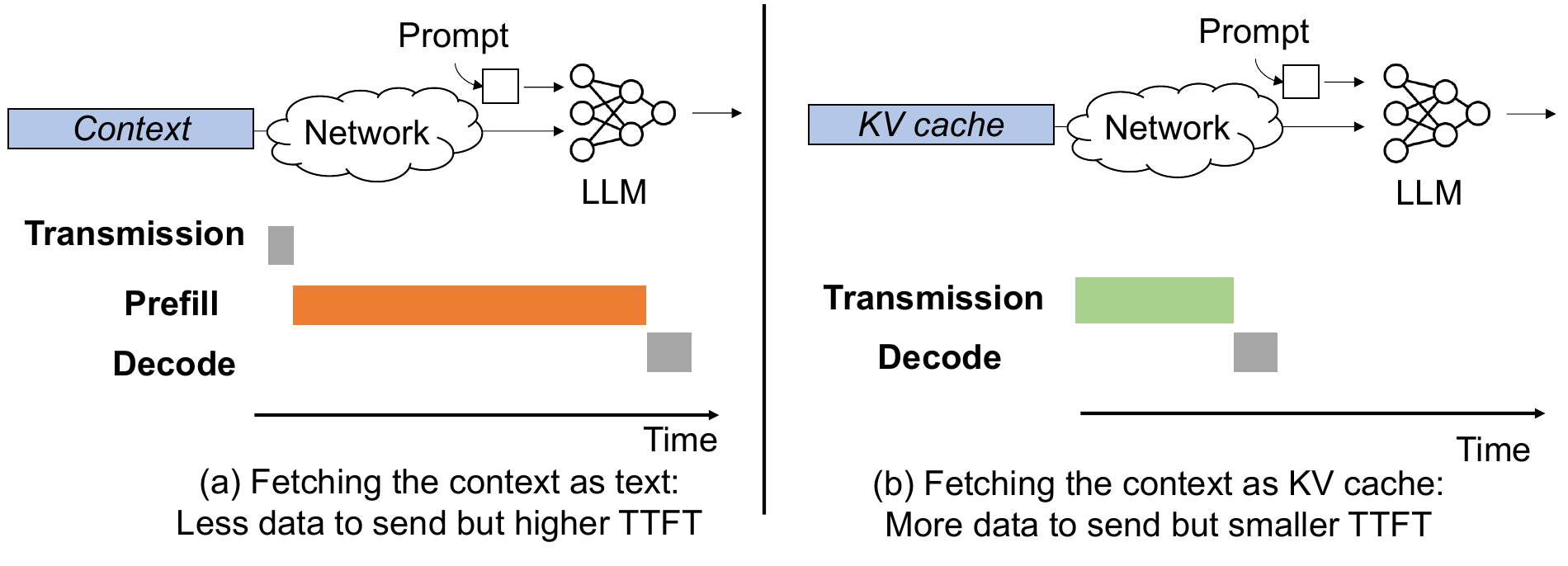}
\tightcaption{Illustration of text recomputation and KV cache reusing. }
\label{fig:input_length}
\end{figure}


\begin{figure*}[]
\centering
\includegraphics[width=0.99\linewidth]{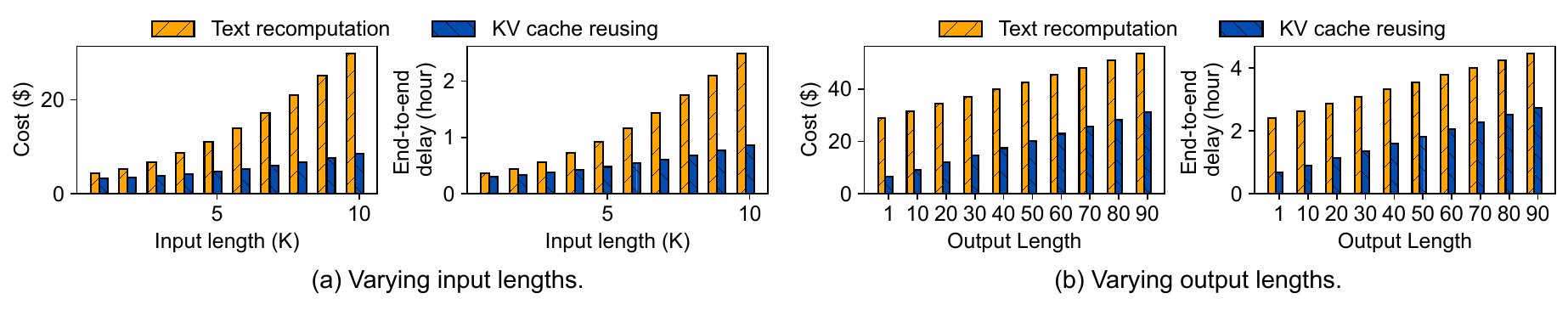}
\tightcaption{The cost and end-to-end delay for Llama-7B  by varying input lengths. }
\label{fig:input_length}
\end{figure*}

\tightsection{Cost Modeling}
\mypara{Analytical Model} In order to compare the operation costs of text recomputation and KV cache reusing, we built an analytical model to measure their costs.
For the first text recomputation pipeline, the inference serving cost (\ie GPU compute) for one request could be estimated by the GPU use time per request times the number of requests using the context within a time period. 
 The breakdown cost can be expressed by the following equation:
\[
C_{text} = C_{GPU} \times N \Bigl[ T_{\text{prefill}} (L_{\text{context}} + L_{\text{prompt}}) + T_{\text{decode}}(L_{\text{output}}) \Bigr],
\]

where $C_{GPU}$ is the cost of the GPU instance per hour, $N$ is number of requests querying the same context, $L_{context}$ is the number of context tokens, $L_{prompt}$ is the token length of the actual prompt,  and $L_{output}$ is the number of generated tokens. $T_{prefill}$ and $T_{decode}$ are time (hour) spent on the computation job given specific token numbers.

For the context reusing pipeline, we separate the cost into three parts: inference, context storage, and transmission cost:
\begin{align}
    C_{KV} = \overbrace{C_{GPU} \times \Bigl\{ N \Bigl[ T_{\text{decode}}(L_{\text{output}}) + T_{\text{prefill}}(L_{\text{prompt}})\Bigr] + T_{\text{prefill}} \Bigr\}  }^{\substack{\text{\sf \footnotesize {GPU inference cost}} }} \nonumber\\ +
    \overbrace{C_{\text{storage}}\times S_{storage}(L_{\text{context}}) \times T}^{\substack{\text{\sf \footnotesize{Storage cost}}}}\nonumber\\ +  \overbrace{C_{\text{transmission}}(S_{storage}(L_{\text{context}}), SLO)}^{\substack{\text{\sf \footnotesize{Transmission cost}}}}\nonumber,
\end{align}

where $C_{\text{storage}}$ is the storage cost per hour$\times$GB, $T$ is the period of serving, $S_{\text{storage}}$ is the size of the KV cache, $SLO$ is the service level objective for time to first token, and  $C_{\text{storage}}$ is the cost per hour to maintain the transmission link that satisfies $SLO$ for loading the KV cache. 

\mypara{Insights} By analyzing the model with real pricing data from AWS, we find that Reusing KV cache could be more economical as long as the context is reused more than once per hour. And that the storage cost is usually only a minimal portion of the total cost. 
Take Llama-7B model as an example. A context of length 10K-token requires a storage space of 5.2~GB.
With Amazon Elastic Block Storage(EBS) that could be directly mounted onto instance, the highest tier storage with 4GB/S throughput, called io2, costs 0.125 per GB $\times$ month~\cite{ebs} Since our IO operations are not frequent, we do not need to pay much extra provisioned IOPS fee.  
Within the period of one hour LLM serving. The total storage and transmission cost spent this hour for this 10K-token context is \$ $8.8e^{-4}$\footnote{io2 storage cost per hour $\times size =0.00017 * 5.2 = 0.00088$ }. 
On the other hand, the GPU inference cost for prefilling is around \$6$e^{-3}$ for a single request \footnote{Single V100 GPU cost per hour $\times \frac{1h}{3600s} \times \T_{prefill}(10K) =$  \$3 $\times \frac{1}{3600} \times 7 = 0.0058$ }. This is already more than 7 times larger than the fixed cost of storage during this hour. And if more requests reuse this piece of context in this hour, this fixed cost in reusing KV cache will be negligible in total cost. We can simplify the ratio between $C_{\text{text}}$ and $C_{\text{KV}}$:

    $\frac{C_{\text{text}}}{C_{\text{KV}}} \approx 1 + \frac{N-1}{N} \frac{T_{\text{prefill}}(L_{\text{context}})}{T_{\text{decode}}(L_{\text{output}}) + T_{\text{prefill}}(L_{\text{prompt}})}$.
    This demonstrates that when context is long while the prompt and output are short, such as in many article Q\&A tasks~\cite{joshi2017triviaqa}, the savings of loading KV cache from storage could be significant.

\tightsection{Preliminary Results}

\mypara{Evaluation setup} In the following experiment, we run Llama-7B model on an Amazon's EC2 p3.8xlarge instance (4 V100 GPUs). 
We apply HuggingFace's model parallelism to run inference on four GPUs. 
The dataset is TriviaQA~\cite{joshi2017triviaqa}, which contains 200 pieces of different contexts. 
We assume each context is used five times and vary two workload parameters, the input lengths and output lengths.

\mypara{Takeaways} 
\begin{packeditemize}
\item When varying the input length from 1--10K, as shown in Figure~\ref{fig:input_length}(a), using KV cache reuse can significantly lower both the end-to-end delay (by  1.1 to 2.9 $\times$) and the cost of using cloud services (by about 1.3 to 3.6 $\times$). 
The benefits of KV cache reuse become more significant as the length of the input increases. 
For shorter inputs (like 1,000 units), the time saved by skipping the prefilling delay does not compromise the delay it takes to load the KV cache.
    \item In Figure~\ref{fig:input_length}(b) where the output lengths are varied from 1 to 100 tokens, the delay saving ranges from 1.6--3.5$\times$ and the cost saving ranges from 1.7--4.5$\times$. 
    We observe that the longer the output is, the less saving KV cache reusing can bring. 
    The insight is that KV cache reusing is only able to save delay for the first token's generation, thus, the longer the output is, the more likely the saving will be amortized. 
\end{packeditemize}


\bibliographystyle{plain}
\bibliography{citations} 


\clearpage
\end{document}